\documentclass[prl,twocolumn,showpacs]{revtex4}
\usepackage{graphicx}
\usepackage{bm}
\usepackage{dcolumn}
\usepackage{bm}
\begin{document}
\title{Cutoff-independent regularization of four-fermion interactions for
color superconductivity}
\author{R.L.S. Farias$^1$, G. Dallabona$^1$, G. Krein$^1$, and O.A. Battistel%
$^2$}
\affiliation{$^1$Instituto de F\'{\i}sica Te\'orica, Universidade Estadual 
Paulista, Rua Pamplona 145, 01405-900 S\~ao Paulo, SP, Brazil \\
$^2$Departamento de F\'{\i}sica, Universidade Federal de Santa
Maria, 97119-900 Santa Maria, RS, Brazil}

\begin{abstract}
We implement a cutoff-independent regularization of four-fermion
interactions to calculate the color-superconducting gap parameter in quark
matter. The traditional cutoff regularization has difficulties for chemical
potentials $\mu$ of the order of the cutoff $\Lambda$, predicting in
particular a vanishing gap at $\mu \sim \Lambda$. The proposed
cutoff-independent regularization predicts a finite gap at high densities
and indicates a smooth matching with the weak coupling QCD prediction for
the gap at asymptotically high densities.
\end{abstract}

 \maketitle

The study of the properties of high density quark matter has attracted great
interest recently - for reviews and extensive lists of references see 
Refs~\cite{reviews}. The earlier studies~\cite{bailin} on this subject found 
energy gaps of the order of a few MeV's. Since gaps of this order are much 
too small to have observable consequences, not much attention was given to 
the subject until recently, when it was shown~\cite{arw,rssv} within the 
context of instanton-motivated four-fermion interactions that gaps of the 
order of $100$~MeV could be obtained. The possibility of gaps of this order 
were corroborated by subsequent study~\cite{son} using weak coupling
renormalization group techniques for QCD. The result of Ref.\cite{son} for
the two-flavor spin-$0$ superconducting gap $\Delta$ can be written as
\begin{equation}
\Delta \sim \frac{\mu}{g^5} \exp\left( - \frac{3\pi^2}{\sqrt 2 g}\right),
\label{gap_son}
\end{equation}
where $\mu$ is the chemical potential and $g = g(\mu)$ is the QCD coupling
constant. This result is clearly nonperturbative, but was derived assuming
weak coupling, an assumption likely to be valid only at very high densities.
Although inapplicable for densities typically found in the interiors of
neutron stars, it seems to be a sound prediction for the color
superconducting gap at asymptotically high baryon number densities. In
particular, using the one-loop running of $g(\mu)$ with $\mu$, 
Eq.~(\ref{gap_son}) predicts that $\Delta(\mu)$ is an increasing function 
of $\mu$. Elaborations and corrections~\cite{corrs} to Eq.~(\ref{gap_son}) 
do not change this behavior.

In view of the inapplicability of weak coupling techniques at densities of
phenomenological interest and the fact that nonperturbative lattice
techniques are not yet sufficiently developed to deal with such problems,
the use of phenomenological models seem to be necessary to make progress in
the field. In this context, models with nonrenormalizable four-fermion
interactions have been extensively used to study different aspects of
dynamical chiral symmetry breaking (D$\chi$SB) and high density quark 
matter~\cite{NJLmatter}. However, four-fermion models at the one loop level 
predict vanishing superconducting gaps at high densities, a feature that is 
caused by the use of a regularizing momentum cutoff $\Lambda$ of the 
divergent one-loop integrals~\cite{casalbuoni, ja, skp}. Since the phenomenon 
of superconductivity involves momenta of the order of the Fermi momentum $k_F$,
for baryon number densities such that $k_F \sim \Lambda$ the cutoff
regularization becomes clearly inadequate and the vanishing of the gap at
high densities might not be a physical feature of the problem. Although the
QCD prediction of Eq.~(\ref{gap_son}) of a nonzero gap is valid only at very
high densities, there seems no physical motivation for expecting vanishing
gaps at moderately high densities. Of course, in principle one cannot exclude 
the possibility that the spin-$0$ gap (considered here) indeed vanishes at 
intermediate values of the chemical potential and then is again different 
from zero at much higher densities, as predicted by QCD - but such a question
cannot be answered without a detailed nonperturbative calculation within
QCD. 

The aim of the present paper is to present an alternative to the
cutoff regularization of point-like four-fermion interactions.
Specifically, the superconducting one-loop integrals are reorganized
through mathematical identities in a way finite integrals become
separated from $\mu$-independent divergent integrals. The finite
integrals are integrated without imposing any restriction to the
integration momenta and the divergent integrals are related to
physical quantities at the D$\chi$SB scale in vacuum. As a result,
one obtains a superconducting gap that is nonvanishing at high
baryon number densities. We will keep the discussion very general,
no specific four-fermion model will be used to explain our
procedure; commitment to a specific model will be made only when
presenting numerical results.

Invariably, at the one-loop level point-like four-fermion interactions of
massless quarks lead to a gap equation for the superconducting gap $\Delta$
of the form
\begin{eqnarray}
1 = \lambda \, G \, i \!\int\frac{d^4k}{(2\pi)^4}\Biggl[ \frac{1}
{k_{0}^{2}-\left(k+\mu \right) ^{2}-\Delta ^{2}} + (\mu \rightarrow - \mu)
\Biggr],  \label{gap_eq_4D}
\end{eqnarray}
where $\lambda$ is a numerical coefficient that depends on the Lorentz,
color and flavor structure of the interaction, and $G$ is the strength of
the four-fermion interaction. Obviously, the above integral is ultraviolet
divergent and a regularization scheme must be specified in order to proceed.
The traditional way to regularize the integral of Eq.~(\ref{gap_eq_4D}) is
via a three- or four-dimensional sharp cutoff, or via a form factor. In view
of the nonrenormalizable nature of the interaction, the regularization
becomes part of the model. Using a three-dimensional sharp cutoff, the
self-consistent solution for $\Delta = \Delta(\mu)$ leads to the result that
$\Delta = 0$ for $\mu \sim \Lambda$ -- the use of a form factor instead
leads to the same qualitative behavior.

The regularization method we advocate here, proposed in Ref.~\cite{thesis}
and used in different contexts~\cite{implicit_ref}, avoids the use
of an explicit regulator and the calculation of any divergent integrals.
Specifically, instead of introducing a cutoff in the integrals of Eq.~(\ref%
{gap_eq_4D}), the integrands are manipulated in a way divergences are
isolated in well-known, $\mu$-independent one-loop divergent integrals that
can be related to divergent integrals of the problem of D$\chi$SB in vacuum.
Finite integrals are integrated without imposing any restriction to their
integrands and the remaining divergent integrals fitted to physical
quantities at the D$\chi$SB scale in vacuum.

This is done in the following way. In order to safely manipulate the
divergent integrals, we initially assume the integral in 
Eq.~(\ref{gap_eq_4D}) regularized by some regulating function 
$f(k^{2}/{\Lambda ^{2}})$, were $\Lambda$ is the regularization parameter. 
We rewrite the gap equation as
\begin{equation}
1 = \lambda \, G \, i \int_{\Lambda }\frac{d^{4}k}{(2\pi )^4} 
\Biggl[\frac{1}{k^2_0 - \left( k + \mu \right)^2 - \Delta^2} + 
(\mu \rightarrow -\mu) \Biggr],
\label{gap_eq_4D_reg}
\end{equation}
where $\int_{\Lambda }$ indicates that the integral is regularized.
The next step is the reorganization of the integrands as mentioned
above. This goal can be achieved by using the identity
\begin{eqnarray}
&&\frac{1}{k_{0}^{2}-\left( k\pm \mu \right) ^{2}-\Delta ^{2}} =
\frac{1}{k_{0}^{2}-k^{2}-\Delta ^{2}} \nonumber \\
&&\hspace{0.75cm}-\frac{\left( \mp \,2k\mu -\mu^{2}\right)} {\left(
k_{0}^{2}-k^{2}-\Delta ^{2}\right) ^{2}} + \frac{\left( \mp \,2k\mu
-\mu ^{2}\right) ^{2}}{\left(k_{0}^{2}-k^{2} - \Delta ^{2}\right)^3}
\nonumber \\
&&\hspace{0.75cm}- \frac{\left( \mp \,2k\mu -\mu ^{2}\right)
^{3}}{\left( k_{0}^{2}-k^{2}-\Delta ^{2}\right) ^{3}\left[
k_{0}^{2}-\left( k\pm \mu \right) ^{2}-\Delta ^{2}\right] },
\label{ident}
\end{eqnarray}
which can be obtained after using three times in succession the identity
\begin{eqnarray}
&&\frac{1}{k_{0}^{2}-\left( k\pm \mu \right) ^{2}-\Delta ^{2}} =\frac{1}{%
k_{0}^{2}-k^{2}-\Delta ^{2}}  \nonumber \\
&&\hspace{1.0cm}+\frac{\mu ^{2}\pm 2k\mu }{\left(
k_{0}^{2}-k^{2}-\Delta ^{2}\right) \left[ k_{0}^{2}-\left( k\pm \mu
\right) ^{2}-\Delta ^{2}\right] }. \label{identidade}
\end{eqnarray}%
This identity corresponds to a subtraction of the original
expression at the point $\mu =0$, akin to what is done with
nonconvergent integrals in dispersion relations. When the identity
of Eq.~(\ref{ident}) is substituted into Eq.~(\ref{gap_eq_4D_reg}),
one can rewrite the gap equation in the form
\begin{eqnarray}
1 &=& 8\lambda G\Bigl\{ 2\left[ iI_{quad}\left( \Delta ^{2}\right)
\right] -4\mu ^{2}\left[ iI_{log}\left( \Delta ^{2}\right) \right]
\nonumber \\ &+& I_{fin}(\Delta ^{2},\mu ) + I_{fin}(\Delta
^{2},-\mu )\Bigr\} ,
\end{eqnarray}
where
\begin{eqnarray}
I_{quad}(\Delta ^{2}) &=& \int_{\Lambda }\frac{d^{4}k}{(2\pi )^{4}}\frac{1}{%
k_{0}^{2}-k^{2}-\Delta ^{2}}, \label{Iquad}\\
I_{log}(\Delta ^{2}) &=& \int_{\Lambda }\frac{d^{4}k}{(2\pi )^{4}}\frac{1}{%
(k_{0}^{2}-k^{2}-\Delta ^{2})^{2}}, \label{Ilog}\\
I_{fin}(\Delta ^{2},\mu ) &=& i\int \frac{d^{4}k}{(2\pi )^{4}}
\frac{1}{(k_{0}^{2}-k^{2}-\Delta ^{2})^{3}} \nonumber \\
&&\hspace{-1.75cm}\times\Biggl[2\mu ^{2}(\mu ^{2}-4\Delta^{2}) +
\frac{(\mu ^{2}+2k\mu )^{3}}{(k_{0}^{2}-(k+\mu )^{2}-\Delta
^{2})}\Biggr]. \label{Ifin}
\end{eqnarray}
In the absence of the regularizing function, $I_{quad}(\Delta ^{2})$ and $%
I_{log}(\Delta ^{2})$ are divergent, but $I_{fin}(\Delta ^{2},\mu )$ is
finite. The divergent integrals are of the same form as those appearing in
the problem of D$\chi $SB in vacuum, with the difference that the mass scale
appearing in the superconductivity integrals is $\Delta $ while in the D$%
\chi $SB integrals the scale is the constituent quark mass $M$. But one can
relate the integrals at different mass scales by making use of the following
scaling properties~\cite{thesis,implicit_ref}
\begin{eqnarray}
I_{quad}(\Delta ^{2}) &=& I_{quad}(M^{2})+(\Delta ^{2}-M^{2})I_{log}(M^{2})
 \nonumber\\
&+&\frac{i}{(4\pi )^{2}}\left[ \Delta ^{2}-M^{2}-\Delta ^{2}\ln \left( \frac{%
\Delta ^{2}}{M^{2}}\right) \right]\label{scale_iquad} ,\\
I_{log}(\Delta ^{2}) &=&I_{log}(M^{2})-\frac{i}{(4\pi )^{2}}\ln \left( \frac{%
\Delta ^{2}}{M^{2}}\right) .  \label{scale_ilog}
\end{eqnarray}
These can be obtained by integrating the relations
\begin{equation}
\frac{\partial I_{quad}\left( \mu ^{2}\right) }{\partial \mu ^{2}} =
I_{log}\left( \mu ^{2}\right) ,\hspace{0.5cm} \frac{\partial
I_{log}\left( \mu ^{2}\right) }{\partial \mu ^{2}}=\frac{-i}{16\pi
^{2}\mu ^{2}},  \label{rel1}
\end{equation}
between the two mass scales $M^{2}$ and $\Delta ^{2}$. The
expressions in Eq.~(\ref{rel1}) can be proven by direct
differentiation of the integrals in Eqs.~(\ref{Iquad}) and
(\ref{Ilog}). In integrating these equations, finite integrals are
integrated without regularization. The relations of
Eqs.~(\ref{scale_iquad}) and (\ref{scale_ilog}) allow us to
normalize the divergent integrals $I_{quad}\left( \Delta ^{2}\right)
$ and $I_{log}\left( \Delta ^{2}\right) $ to chiral observables at
the scale of the constituent quark mass $M$. This is so because
$I_{quad}\left( M^{2}\right) $ and $I_{log}\left( M^{2}\right)$ in
this class of models are related respectively to the chiral
condensate $\langle \bar{\psi}\psi \rangle $ and to the pion decay
constant $f_{\pi }$ (in the chiral limit) by
\begin{equation}
iI_{quad}(M^{2})=\frac{-\langle \bar{\psi}\psi \rangle
}{12M},\hspace{0.5cm} iI_{log}(M^{2})=-\frac{f_{\pi }^{2}}{12M^{2}}.
\label{fits}
\end{equation}
Therefore, making use of the relations of Eqs.~(\ref{scale_iquad}) and (\ref%
{scale_ilog}), one can eliminate $I_{quad}(\Delta ^{2})$ and $I_{log}(\Delta
^{2})$ in the gap equation for $\Delta $ in favor of $\langle \bar{\psi}\psi
\rangle $ and $f_{\pi }$, obtaining
\begin{eqnarray}
1 &=& \lambda \,G\Bigl[2I_{1}(M^{2},\Delta ^{2})-4\mu
^{2}I_{2}(M^{2},\Delta ^{2}) \nonumber \\
&+& I_{fin}(\Delta ^{2},\mu ) + I_{fin}(\Delta ^{2},-\mu )\Bigr],
\label{gap-fin}
\end{eqnarray}
where
\begin{eqnarray}
\hspace{-0.35cm}I_{1}(M^{2},\Delta ^{2}) &=&-\frac{\langle \bar{\psi}\psi 
\rangle }{12M} - (\Delta^2 - M^2)\frac{f_{\pi }^{2}}{12M^{2}}  \nonumber \\
&-&\frac{1}{(4\pi )^{2}}\left[ \Delta^2-M^2-\Delta^2 \ln \left( 
\frac{\Delta ^{2}}{M^{2}}\right) \right], \\
\hspace{-0.35cm}I_{2}(M^{2},\Delta ^{2}) &=&\frac{f_{\pi }^{2}}{12M^{2}}
- \frac{1}{(4\pi )^2}\ln \left( \frac{\Delta ^{2}}{M^{2}}\right) .
\end{eqnarray}
In this way, we have eliminated the divergences without specifying any
regularization function. The use of $f(k^{2}/{\Lambda ^{2}})$ is simply a
matter of formality, in the sense that once the integrals are regularized,
they can be freely manipulated. In a renormalizable theory, one would simply
do the same manipulations but the remaining divergences would be eliminated
by counterterms added to the original Lagrangian (or Hamiltonian). The basic
divergent integrals could also be written in terms of an arbitrary mass
scale $\hat{m}$, which can be used to relate observables at different mass
scales by using the standard renormalization group techniques. The entire
regularization-renormalization process can be done without specifying any
regularization function and without calculating any divergent integral. In
the present case with nonrenormalizable interactions, the divergences are
simply fitted to physical quantities at the chiral symmetry mass scale $M$.
Note that instead of referring to a regularizing function $f(k^{2}/{\Lambda
^{2}})$ parameterized in terms of a momentum scale $\Lambda $, one could
have simply used dimensional regularization, fit the divergent integrals to $%
\langle \bar{\psi}\psi \rangle $ and $f_{\pi }$ as above, and evaluate
finite integrals at the physical dimension.

\begin{figure}[ht]
\includegraphics[scale=1.0]{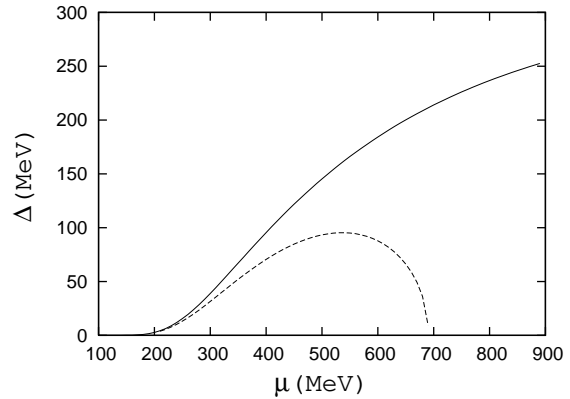}
\caption{Superconducting gap $\Delta $ as a function of the quark chemical
potential. The solid line is the solution of Eq.~(\protect\ref{gap-fin}) and
the dashed line is the result using a cutoff in $I_{\mathrm{fin}}$. }
\label{fig:impl}
\end{figure}

So far, the discussion has been completely general, valid for any point-like
four-fermion interaction. In order to see the physical consequences of our
proposed method, we solve the gap equation for a specific four-fermion
model. For illustrative purposes, we use the simple chirally symmetric $%
SU(2) $ Nambu--Jona-Lasinio model with scalar-plus-pseudoscalar four-fermion
interactions~\cite{skp}. For such a model, $\lambda = 8 N_f$, where $N_f = 2$
is the number of flavors. The model needs only two parameters as input, one
is the strength of the interaction and the other is a regularization
parameter which in the present approach can be taken to be one of the
divergent integrals. Using a standard value for the strength of the
interaction, $G~=~3.1$~GeV$^{-2}$, and $\langle \bar{\psi}\psi \rangle =
(-~250\,\mathrm{{MeV})^{3}}$ it is possible to obtain
approximately $M=313$~MeV and $f_{\pi}~=~ 93$~MeV. With these values, one
obtains the solid line in Fig.~\ref{fig:impl}. Clearly, this results shows
that the superconducting gap $\Delta$ is an increasing function of the
chemical potential $\mu$. Although we have not plotted the result for larger
values of $\mu$, the curve actually keeps increasing very slowly as a
logarithmic function, and for $\mu \sim 3$~GeV the gap is $\Delta \simeq 300$%
~MeV. The basic reason for the nonvanishing of $\Delta$ at high values of $%
\mu$ is that no momentum cutoff is enforced on the finite integral and
therefore there is enough phase space at large chemical potentials for
allowing larger gaps. To substantiate this we have imposed a cutoff $\Lambda
= 650$~MeV on $I_{fin}$ and obtained the dashed curve in Fig.~\ref{fig:impl}%
. The result is clear, the gap vanishes for chemical potentials of the order
of the cutoff $\mu \sim \Lambda$. Note that for consistency one should have
used the same cutoff in the finite integrals that appear in the derivation
Eqs.~(\ref{scale_iquad}) and (\ref{scale_ilog}), but this would simply
worsen the situation, in the sense that $\Delta$ would be zero for $\mu <
700 $~MeV. We repeated the calculation using other types of point-like
four-fermion interactions and obtained qualitatively similar results.

The same scheme can be used at finite temperatures~\cite{run-us},
the gap equation would be given by an expression similar to
Eq.~(\ref{gap-fin}), with the difference that $I_{fin}(\Delta^2,
\mu)$ would include a term with the Fermi-Dirac distributions for
quarks and anti-quarks. In connection to this, it is worth
mentioning that the integration of finite integrals without imposing
a cutoff has been advocated in Refs.~\cite{ruivo} in the context of
D$\chi$SB, where the finite-temperature integrals with the
Fermi-Dirac distributions are freely integrated, while the divergent
integrals are still evaluated with a cutoff $\Lambda$. The crucial
difference with the present approach is the use of the scaling
relations of Eqs.~(\ref{scale_iquad}) and (\ref{scale_ilog}), by
which one can always relate the divergent integrals to vacuum
quantities without commitment to a specific regularization. Detailed
finite temperature results using the scheme proposed here will be
presented elsewhere~\cite{run-us}.

Before concluding, one should notice that although we have not
imposed a cutoff on finite integrals, there is still, of course, an
implicit regularization scale dependence in the model through the
fitting $I_{quad}$ and $I_{log}$ to vacuum physical quantities
$\langle \bar\psi\psi\rangle$ and $f_\pi$ as in Eq.~(\ref{fits}).
This dependence reflects the fact that the physical quantities (and
the coupling~$G$) are fitted at the implicit regularization scale
$\Lambda$. Equivalently, had we used explicit dimensional
regularization, a mass scale would enter the problem to match the
physical dimensions of the integrals and this scale would also be
implicit in the fitting of the divergent integrals. This scale sets
a limit on the chemical potential $\mu$ for which the fitting makes
physical sense. Therefore, to use a four-fermion model at very high
densities one must devise a scheme to extend the model beyond the
vacuum scale. This actually can be done using the method proposed in
Ref.~\cite{casalbuoni} by making the coupling run with $\Lambda$,
i.e. by making $G =G(\Lambda)$, and postulating a $\mu$ dependence
of $\Lambda$. In Ref.~\cite{casalbuoni} this was done by demanding
$\Lambda$-independence of $f_\pi$ in a cutoff regularization scheme.
In the present approach such a method can be applied using the
scaling properties of the divergent integrals $I_{quad}$ and
$I_{log}$. A detailed discussion on this will be presented in a
separate publication~\cite{run-us}. In this context, it would
also be interesting to implement the method discussed here in
instanton-motivated interactions, where the cutoff has a physical
origin and has a density dependence such that the associated form factor 
peaks around the Fermi momentum with a width of the order of the inverse 
size of the instanton~\cite{inst-sch}. Since the instanton effects at 
large density are suppressed because the effective coupling becomes 
small, the present method might be useful for handling a possible 
smooth matching to perturbative QCD.

In summary, we have shown that using a cutoff-independent regularization of
nonrenormalizable point-like four-fermion interactions at the one-loop level
one obtains nonvanishing superconducting gaps at high densities in quark
matter. The result also indicates a smooth matching without abrupt
discontinuities with the weak coupling QCD prediction for the gap~\cite%
{son,corrs}. Although this QCD prediction of a nonzero gap
is valid only at asymptotically high quark densities, it seems nevertheless
reasonable to assume that the vanishing of the gap at high densities in
four-fermion models is a nonphysical artifact of the cutoff regularization.
Admittedly, there is implicit an assumption that four-fermion models can be 
used at energies higher than the chiral symmetry breaking scale. This can be 
naturally implemented by making regularization parameters density dependent,
as physically motivated in Ref.~\cite{casalbuoni}. In this sense, the 
present approach opens new opportunities for applications of four-fermion 
models for studying properties of high density quark matter without facing 
the difficulties imposed by a cutoff. Obviously, the technique presented here 
is applicable in a wider context where nonrenormalizable interactions are 
used for modelling physical systems, such as atomic condensates, 
for example.

\acknowledgments

The authors thank Thomas Sch\"afer and Qun Wang for useful correspondence.
This work was partially supported by CNPq and FAPESP (Brazilian agencies).

\end{document}